\documentclass[aps,prd,superscriptaddress,twocolumn,floatfix,nofootinbib]{revtex4-1}
\usepackage{graphicx}
\graphicspath{{images/}}
\usepackage{amsmath}
\usepackage{amsfonts}
\usepackage{amssymb}
\usepackage{array}
\usepackage{booktabs}
\usepackage{epsfig}
\usepackage{epstopdf}
\usepackage{bm}
\usepackage[normalem]{ulem}  
\usepackage{color} 

\begin{document}


\title{The negative-parity spin-1/2 $\Lambda$ baryon spectrum from lattice QCD and effective theory}



\author{Rafael Pavao}
\affiliation{Instituto de F\'isica Corpuscular (IFIC),
Centro Mixto CSIC-Universidad de Valencia,
Institutos de Investigac\'ion de Paterna,
Apartado 22085, E-46071 Valencia, Spain}

\author{Philipp Gubler}
\affiliation{Advanced Science Research Center, Japan Atomic
Energy Agency, Tokai, Ibaraki 319-1195, Japan}

\author{Pedro Fernandez-Soler}
\affiliation{Instituto de F\'isica Corpuscular (IFIC),
Centro Mixto CSIC-Universidad de Valencia,
Institutos de Investigac\'ion de Paterna,
Apartado 22085, E-46071 Valencia, Spain}

\author{Juan Nieves}
\affiliation{Instituto de F\'isica Corpuscular (IFIC),
Centro Mixto CSIC-Universidad de Valencia,
Institutos de Investigac\'ion de Paterna,
Apartado 22085, E-46071 Valencia, Spain}

\author{Makoto Oka}
\affiliation{Advanced Science Research Center, Japan Atomic
Energy Agency, Tokai, Ibaraki 319-1195, Japan}

\author{Toru T. Takahashi}
\affiliation{National Institute of Technology, Gunma College, Gunma 371-8530, Japan}


\date{\today}

\begin{abstract}
The spectrum of the negative-parity spin-1/2 $\Lambda$ baryons 
is studied using lattice QCD and hadronic 
effective theory in a unitarized coupled-channel framework. A direct comparison between the two 
approaches is possible by considering the hadronic effective theory in a finite volume and 
with hadron masses and mesonic decay constants that correspond to the situation studied 
on the lattice. 
Comparing the energy level spectrum and $SU(3)$ flavor decompositions of the individual states, 
it is found that the lowest two states extracted from lattice QCD can be associated with 
one of the two $\Lambda(1405)$-poles and the $\Lambda(1670)$ resonance.  
The quark mass dependences of these two lattice QCD levels are in good 
agreement with their effective theory counterparts. 
However, as current lattice QCD studies still rely on three-quark operators to generate 
the physical states, clear signals corresponding to the meson-baryon scattering states, 
that appear in the finite volume effective theory calculation, are not yet seen.   
\end{abstract}


\maketitle

\section{Introduction \label{sec:Intro}}
The spectrum of
negative-parity $\Lambda$ baryons has a rich structure, that has been studied 
intensively during the last decades. The lowest spin-1/2 member, the $\Lambda(1405)$, whose existence was 
already anticipated in the late fifties by Dalitz and Tuan \cite{Dalitz:1959dn,Dalitz:1960du}, has especially attracted much 
interest because it can naturally be interpreted as a $\bar{K}N$ and $\pi \Sigma$ molecule, rather than as an ordinary three-quark 
state \cite{Dalitz:1967fp}. 
This means that its properties can provide valuable information about the strength of the phenomenologically 
important $\bar{K} N$ interaction \cite{Akaishi:2002bg}. 
Furthermore, studies based on the chiral unitary approach (CUA) \cite{Kaiser:1995eg,Oset:1997it} 
have found that the $\Lambda(1405)$ is not composed of only one single pole, 
but rather of two poles \cite{Oller:2000fj,GarciaRecio:2002td,Jido:2003cb,GarciaRecio:2003ks,Hyodo:2003qa,Ikeda:2011pi,Guo:2012vv,Mai:2012dt}, 
which is a natural consequence 
of the chiral group dynamics of the lowest order Weinberg-Tomozawa (WT) meson-baryon interaction. 
A similar conclusion was also obtained in an earlier calculation based on the cloudy bag model in Ref.\,\cite{Fink:1989uk}.

Reproducing the properties of the negative parity $\Lambda$ baryon spectrum in a first-principle lattice QCD calculation 
has, on the other hand, proven to be a difficult and computationally demanding task. The early calculations 
\cite{Leinweber:1990dv,Melnitchouk:2002eg,Nemoto:2003ft,Burch:2006cc,Ishii:2007ym}  
had to rely on the quenched approximation (that is, ignoring processes that require virtual quark-antiquark pairs) and on rather 
heavy quark masses, and were therefore still relatively far from reality. 
As simulations including dynamical quarks closer to the physical point became increasingly available, 
the calculations further improved \cite{Takahashi:2009bu}, until the extraction of a 
$\Lambda(1405)$ signal was finally reported in Ref.\,\cite{Menadue:2011pd} and further 
studied in Refs.\cite{Engel:2013ig,Menadue:2013xqa,Hall:2014uca,Hall:2014gqa,Liu:2016wxq,Gubler:2016viv,Menadue:2018abc}. 
By measuring the electromagnetic response 
of the extracted signal, even some first evidence for the molecular nature of the $\Lambda(1405)$ 
was provided in Ref.\,\cite{Hall:2014uca}. The authors of Ref.\,\cite{Gubler:2016viv} furthermore studied 
the $SU(3)$ flavor decomposition of the $\Lambda(1405)$ and investigated the effect of 
the strange quark mass by gradually increasing its value from the strange quark mass to 
the charm quark mass domain. 

In the present work, we employ the CUA and its extension to finite volume for a
detailed comparison between the lattice data of Refs.\,\cite{Gubler:2016viv,Menadue:2018abc} and the effective 
theory results for the low-lying negative parity $\Lambda$ baryon spectrum. 
Similar studies have already been performed for instance 
in Refs.\,\cite{MartinezTorres:2012yi,Molina:2015uqp,Liu:2016wxq,Tsuchida:2017gpb}. 
One focus of this work is the $SU(3)$ flavor structure of the obtained $\Lambda$ states, which 
has so far only been discussed in Refs.\,\cite{Menadue:2011pd,Menadue:2013xqa,Gubler:2016viv} 
on the lattice, but can provide important information 
for identifying an extracted lattice QCD energy level with a corresponding effective 
theory state. 
The lowest $\Lambda(1405)$ will, because of its unique properties, be the major target of 
this study. We will, however, also investigate the next resonance, the $\Lambda(1670)$ 
and examine to what degree it can be extracted from the lattice results of Refs.\,\cite{Gubler:2016viv,Menadue:2018abc}. 
We will furthermore have a broad look at the complete energy level spectrum obtained from CUA at finite volume, 
including meson-baryon scattering states, and discuss if and how the different energy 
levels are reflected in the lattice data. 

This work is organized as follows. In Section\,\ref{sec:Form}, we briefly review the formalisms of 
the lattice calculation of Ref.\,\cite{Gubler:2016viv} and the CUA for infinite and finite volumes. 
Section\,\ref{sec:results} contains the comparison between the lattice QCD and CUA results 
for the spectrum and a discussion of the flavor structure of the individual energy levels. 
The paper is summarized and concluded in Section\,\ref{sec:SumCon}.

\section{Formalism \label{sec:Form}}
\subsection{Lattice QCD}
Let us here summarize the methods adopted in Ref.\,\cite{Gubler:2016viv}. 
In this work, 
gauge configurations with 2+1 active flavors were employed, which were originally generated by the 
PACS-CS Collaboration \cite{Aoki:2008sm}. 
A renormalization-group-improved action for the gauge fields and $\mathcal{O}(a)$-improved action 
for quarks were used. The gauge coupling was set to $\beta=1.9$, which translates to a lattice spacing 
of $a = 0.0907\,\mathrm{fm}$ at the physical point \cite{Aoki:2008sm}. The employed lattice size is 
$32^3 \times 64$, leading to a physical extent of $L = 2.90\,\mathrm{fm}$. 
We note that in the study by the Adelaide group \cite{Hall:2014uca}, the same gauge configurations were used. 
In the quark action, the strange quark hopping parameter was fixed to $0.13640$, while for the up and down 
quarks it was set to $0.13700$, $0.13727$, $0.13754$, and $0.13770$, which corresponds to 
pion masses between $700\,\mathrm{MeV}$ and $290\,\mathrm{MeV}$. 

To generate the $\Lambda$ baryon states, three-quark operators of the following general form 
were adopted, 
\begin{equation}
\Lambda_{\mu_1 \mu_2 \mu_3} = 
\frac{1}{\sqrt{2}} \epsilon_{abc} (u^a_{\mu_1} d^b_{\mu_2} - d^a_{\mu_1} u^b_{\mu_2}) s^c_{\mu_3}, 
\label{eq:operators}
\end{equation} 
where $a$, $b$, $c$ ($\mu_1$, $\mu_2$, $\mu_3$) are color (Dirac) indices. 
Different components of the above operator were chosen following Ref.\,\cite{Basak:2005ir}, where it was shown that 
four independent spin-1/2 operators can be constructed for both positive and negative parities. 
For the rest of this section, we will denote them as $\Lambda_{I}$ ($I = 1,\dots,4$) for simplicity of 
notation. 
Among these, three ($I = 1,2,3$) 
can be categorized as flavor-octet operators, while one ($I=4$) belongs to the flavor-singlet 
irreducible representation of $SU(3)$ flavor. 

To extract the lowest lying states of the spectrum and their flavor content, we compute the 
correlator matrix 
\begin{equation}
\mathcal{M}(x,y)_{IJ} = \langle \Lambda_{I}(x) \bar{\Lambda}_{J}(y) \rangle.   
\end{equation}
By inserting two complete sets of energy 
eigentstates $|i\rangle$, $\mathcal{M}(t,0)_{IJ}$ can be rewritten as 
\begin{equation}
\mathcal{M}(t,0)_{IJ} = \sum_{n,m} (C^{\dagger}_{\mathrm{snk}})_{In} \Omega(t)_{nm}(C_{\mathrm{src}})_{mJ},    
\end{equation}
where $C^{\dagger}_{\mathrm{snk}}$ and $C_{\mathrm{src}}$ represent matrix elements 
between the three-quark $\Lambda$ operators and energy eigenstates, $(C^{\dagger}_{\mathrm{snk}})_{In} = \langle \mathrm{vac} |\Lambda_{I} | n \rangle$, 
$(C_{\mathrm{src}})_{mJ} = \langle m | \bar{\Lambda}_{J}| \mathrm{vac} \rangle$. 
The diagonal matrix $\Omega(t)_{nm}$ is related to the (imaginary) time evolution and can be 
given as 
\begin{equation}
\Omega(t)_{nm} = \delta_{nm} e^{-E_n t}, 
\end{equation}
where $E_n$ is the eigenenergy of the state $|n\rangle$. 
The lowest lying eigenenergies and the overlap of the respective states with 
the operators $\Lambda_{I}$ can be obtained (for instance) by computing the 
eigenenergies and eigenvectors of the matrix products $\mathcal{M}^{-1}(t+1) \mathcal{M}(t)$ 
or $\mathcal{M}(t) \mathcal{M}^{-1}(t+1)$. For more details, we refer the 
reader to Ref.\,\cite{Gubler:2016viv}. 
For a numerical estimate of the flavor decomposition of the found states, we use  
$\Psi^{1/2}_{In} \equiv \langle \mathrm{vac} |\Lambda_{I} | n \rangle$ to define 
\begin{align}
g_n^{\bm{1}} &= \frac{|\Psi_{4n}|}{\sum_{I=1}^{4}|\Psi_{In}|}, \label{eq:su3_latt_1} \\
g_n^{\bm{8}} &= \frac{  \sum_{J=1}^{3}|\Psi_{Jn}|  }{\sum_{I=1}^{4}|\Psi_{In}|}, \label{eq:su3_latt_2}
\end{align}
which is normalized such that $g_n^{\bm{1}} + g_n^{\bm{8}} = 1$. \\

\subsection{The chiral unitary approach}
We here give a short overview of the CUA to be used in this work, 
based on Ref.~\cite{GarciaRecio:2003ks}. 
The necessary hadronic building blocks in this approach are 
the $I = 0$ channels $\pi \Sigma$, $\bar{K} N$, $\eta \Lambda$ and $K \Xi$, 
which correspond to the channels 1, 2, 3 and 4 in the notation of this paper, 
together with their coupled-channel chiral interactions. 
With these, 
the two $\Lambda(1405)$ and the $\Lambda(1670)$ states in the meson-baryon scattering amplitude 
could be described \cite{Oset:2001cn,GarciaRecio:2003ks}. 

To obtain such poles, the meson-baryon scattering 
amplitude $T(s)$ ($s$ being the total center-of-mass energy) is unitarized using the Bethe-Salpeter equation (BSE), 
which can schematically be written as 
\begin{equation}
\label{eq:inftyvol}
T = \left[1 - V G \right]^{-1} V. 
\end{equation}
For the $V_{ij}$ kernel, we take 
the WT kernel, lowest order in the chiral expansion, 
\begin{equation}
V^{WT}_{ij}(s) = D_{ij} \frac{2 \sqrt{s} - M_i - M_j}{4 f_i f_j},  
\end{equation}
where $M_i$ and $M_j$ ($f_i$ and $f_j$) are the baryon masses (meson decay constants) of channel $i$ and $j$, respectively. 
The matrix $D_{ij}$ is given as 
\begin{equation}
\label{eq:Dmat}
D=\begin{pmatrix}
-4 & \sqrt{\frac{3}{2}} & 0 & -\sqrt{\frac{3}{2}} \\
\sqrt{\frac{3}{2}} & -3 & -\sqrt{\frac{9}{2}} & 0 \\
0 & -\sqrt{\frac{9}{2}} & 0 & \sqrt{\frac{9}{2}} \\
-\sqrt{\frac{3}{2}} & 0 & \sqrt{\frac{9}{2}} & -3
\end{pmatrix}. 
\end{equation}
In addition, 
$G_i(s)$ is the loop function, which is regularized using a subtraction at an energy scale $\mu$,
\begin{equation}
G_i(s) = \overline{G}_i(s)- \overline{G}_i(\mu^2),
\end{equation}
with 
$\mu = M_{\Lambda}$ \cite{Lutz:2001yb}. 
The ultraviolet (UV) finite part is given as
\begin{widetext}
\begin{equation}
\overline{G}_i(s) = \frac{2 M_i}{16 \pi^2} \left \{ \left[ \frac{M_i^2-m_i^2}{s} - \frac{M_i-m_i}{M_i+m_i} \right] \log \frac{M_i}{m_i}  
+ \frac{2 |\bm{k}_i|}{\sqrt{s}} \left [ \log \frac{1+\sqrt{\frac{s-s_{i +}}{s-s_{i -}}}}{1-\sqrt{\frac{s-s_{i +}}{s-s_{i -}}}} - i \pi \right] \right\}, 
\end{equation}
\end{widetext} 
where $s_{i \pm}$ and $|\bm{k}_i|$ are defined as 
$s_{i \pm} = (M_i \pm m_i)^2$ 
and 
$|\bm{k}_i| = \frac{\lambda^{1/2}(s, M_i, m_i)}{2\sqrt{s}}$ 
with $\lambda(x,y,z) = x^2 + y^2 + z^2 - 2xy - 2xz - 2yz$. 
Through unitarization, the
physical states are generated as poles being located in the complex $s$ plane of $T(s)$, bound states (resonances) in 
the first (second) Riemann sheet (see for instance Sec. IIID of Ref.\,\cite{Nieves:2001wt}). 
The mass and width of the resonances are encoded in the positions of the poles on the complex energy plane.

\subsection{Compositeness \label{sec:Compo}}
Any of the negative parity $\Lambda$ baryon states studied in this work 
can be considered to be composed of either an intrinsic three-quark state, a meson-baryon molecular state 
or generally of a mixture of the two configurations. 
In this section, we discuss how a state can be characterized with the help of the so-called compositeness. 

For a bound/resonance state, one can calculate the relative size of the meson-baryon molecule admixture $| MB_i \rangle$  
by using the compositeness $X_i$, given by \cite{Hyodo:2013nka}
\begin{equation}
\label{eq:comp}
X_i = - \text{Re} \left[\sum_j g_i g_j \frac{\partial G_j}{\partial\sqrt{s}}\Big |_{\sqrt{s} \rightarrow \sqrt{s_P}} \delta_{ij} \right], 
\end{equation}
where the couplings $g_i$ and $g_j$ are calculated from the behavior of the amplitude $T$ close to 
the pole of interest, 
\begin{equation}
T_{ij} \simeq \frac{g_i g_j}{\sqrt{s}-\sqrt{s_P}},
\end{equation}
with $\sqrt{s_P} = M_P - i \Gamma_P/2$, 
$M_R$ and $\Gamma_P$ being the mass and width of the considered state (for a bound state, $\Gamma_P = 0$). 
The three-quark (or any other non-molecular) component of the pole can be determined using
\begin{equation}
Z = -  \text{Re} \left[\sum_j g_i g_j  G_i G_j \frac{\partial V_{ij}}{\partial \sqrt{s}}\Big |_{\sqrt{s} \rightarrow \sqrt{s_P}}\right]. 
\label{eq:Z_def}
\end{equation}
For $X_i$ and $Z$, the normalization condition $\sum_i X_i + Z = 1$ holds. It should, however, be noted that for resonances,  
$X_i$ is generally not guaranteed to satisfy $1 \geq X_i \geq 0$\footnote{At a complex pole, $g_{i(j)}$ and $G_{i(j)}$ in Eqs.\,(\ref{eq:comp}-\ref{eq:Z_def}) are
also complex. We therefore define $X_i$ and $Z$ as the real part
in Eqs.\,(\ref{eq:comp}) and (\ref{eq:Z_def}).} \cite{Aceti:2014ala}. 
This exemplifies the inherent difficulty in unambiguously characterizing and 
classifying resonances as three quark or hadronic molecular states (see Ref.\,\cite{Hyodo:2013nka} for a detailed discussion of this issue). 

One can compute $X_i$ and $Z$ either in the particle or flavor $SU(3)$ basis. 
These bases are connected by relations between the different (isoscalar) eigenstates, 
$|MB_i \rangle = \sum_j c_{ij} | \text{irrep}_j  \rangle$, 
which is given as
\begin{equation}
\begin{pmatrix} |\pi \Sigma \rangle \cr | \bar{K} N \rangle \cr |\eta \Lambda \rangle \cr | K \Xi \rangle \cr\end{pmatrix}
= \sqrt{\frac{1}{40}} \begin{pmatrix} - 1&-  \sqrt{24}&0& \sqrt{15} \cr
- \sqrt{6} &-  \sqrt{4}&\sqrt{20}& -  \sqrt{10}\cr
\sqrt{27}&-  \sqrt{8}&0&- \sqrt{5}\cr
\sqrt{6}&\sqrt{4}&\sqrt{20}&\sqrt{10}\cr
\end{pmatrix}
\begin{pmatrix} |\mathbf{27} \rangle \cr | \mathbf{8_1} \rangle \cr | \mathbf{8_2} \rangle \cr | \mathbf{1}\rangle \cr\end{pmatrix}
\end{equation}
We will make use of both bases in our discussions of the results given in Section\,\ref{sec:results}. 
%

\subsection{Finite volume CUA with non-physical hadron masses}
To study how the states dynamically generated within the CUA could appear in a 
finite-volume lattice simulation, one has to adapt the CUA to the finite volume case. 
For this purpose, we follow the formalism developed in Ref.~\cite{Doring:2011vk}. In a finite box, 
the three-momentum $\bm{q}$ in the loop function is discretized, while the BSE becomes
\begin{equation}
\label{eq:bsfinite}
\tilde{T}(s) = \left[V^{-1}(s) - \tilde{G}(s) \right]^{-1}, 
\end{equation}
with the tilded quantities representing the quantities affected by the discrete behavior of $\bm{q}$: $\bm{q} = 2\pi\bm{n}/L$, with 
$\bm{n} \in \mathbb{Z}^3$. 
At this level, we neglect the dependence of the two-particle irreducible amplitude ($V$) on the finite length box, which would be induced by the left-hand integrations. 
To take into account the most relevant finite-box effects, one in practice has to make the substitution,
\begin{equation}
\label{eq:new_loop}
\tilde{G}_i(s) \rightarrow G_i(s) + \lim_{\Lambda \rightarrow \infty} \left \{ \left( \frac{1}{L^3} \sum_{\vec{n}}^{q < \Lambda} - \int_0^\Lambda \frac{q^2 dq}{2 \pi^2} \right) I_i(q) \right\},
\end{equation}
with $q = |\bm{q}|$ and 
\begin{equation}
I_i (q) = 2 M_i \frac{1}{2 \omega_i(q) E_i(q)} \frac{E_i(q)+\omega_i(q)}{s-(E_i(q)+\omega_i(q))^2 + i \epsilon}, 
\end{equation}
where $E_i(q) = \sqrt{q^2 +M_i^2}$ [$\omega_i(q) = \sqrt{q^2 + m_i^2}$] stands for the baryon [meson] energy. 
Using Eq.~\eqref{eq:bsfinite} with the new loop function of Eq.~\eqref{eq:new_loop}, one can calculate the finite volume energy levels by 
determining the poles of $\tilde{T}$ on the real axis of the complex energy plane. 
Since for a finite spatial box the momenta are not continuous, the scattering states will no longer form a branch cut and an investigation of a second Riemann sheet is no longer possible. 
Resonances will hence manifest themselves only as distortions of the trajectories of the finite volume energy-levels. 

To make a realistic comparison with the lattice QCD results possible, we adopt the hadron masses and decay 
constants that correspond 
to the situation simulated on the lattice. We therefore use in the calculations 
of this work the hadron masses obtained in Ref.\,\cite{Gubler:2016viv}, which we compile here in Table \ref{tab:input_values} together with the physical hadron masses quoted from 
the PDG \cite{Tanabashi:2018oca}. 
For the $\eta$ and $\Xi$ masses that were not computed in Ref.\,\cite{Gubler:2016viv}, we make use of flavor $SU(3)$ relations 
to obtain $m_{\eta} = \sqrt{(4m_K^2 - m_{\pi}^2)/3}$ and $M_{\Xi} = (M_{\Sigma} + 3M_{\Lambda})/2 - M_N$, where the $\eta$ is assumed to be a pure 
$SU(3)$ octet state. 
The meson decay constants depend on the quark masses as well. To take this effect into account, we follow the Appendix of Ref.\,\cite{Molina:2015uqp} 
(see also Ref.\,\cite{Nebreda:2010wv}), and the formulas therein, based on chiral perturbation theory. 
\begin{table*}
\renewcommand{\arraystretch}{1.2}
\centering
\caption{Masses and decay constants employed in the CUA calculation, in units of MeV. For the masses ($m_i$ and $M_i$), 
the first line contains the experimental values taken from \cite{Tanabashi:2018oca}, 
while the others (except for $m_{\eta}$ and $M_{\Xi}$, see text) are extracted from the lattice QCD results 
of Ref.\,\cite{Gubler:2016viv}. For the decay constants ($f_i$), we follow the method presented in the Appendix of Ref.\,\cite{Molina:2015uqp}.}
\begin{tabular}{cccccccc|ccc}  
\toprule
set & $m_{\pi}$ & $m_{K}$ & $m_{\eta}$ & $M_{N}$ & $M_{\Lambda}$ & $M_{\Sigma}$ & $M_{\Xi}$ & $f_{\pi}$ & $f_{K}$ & $f_{\eta}$ \\ \midrule
1 & 138.04 & 495.31 & 547.86 & 938.92 & 1115.68 & 1193.15 & 1318.29 & 92.40 & 112.69 & 121.49 \\
2 & 287.83 & 597.63 & 669.78 & 1126.96 & 1290.12 & 1346.69 & 1481.58 & 102.12 & 118.15 & 126.60 \\
3 & 412.27 & 641.36 & 701.29 & 1218.33 & 1398.90 & 1385.85 & 1572.95 & 109.97 & 120.67 & 126.16 \\
4 & 573.27 & 717.94 & 760.07 & 1411.96 & 1512.03 & 1533.79 & 1622.99 & 117.20 & 122.66 & 125.32 \\
5 & 700.54 & 788.00 & 815.07 & 1555.55 & 1657.80 & 1633.87 & 1748.09 & 120.34 & 123.01 & 124.21 \\
\bottomrule
\label{tab:input_values}
\end{tabular}
\end{table*}

\section{Results \label{sec:results}}
\subsection{Spectrum}
\begin{figure}
\centering
  \includegraphics[scale=0.5,bb=0 0 396 278] {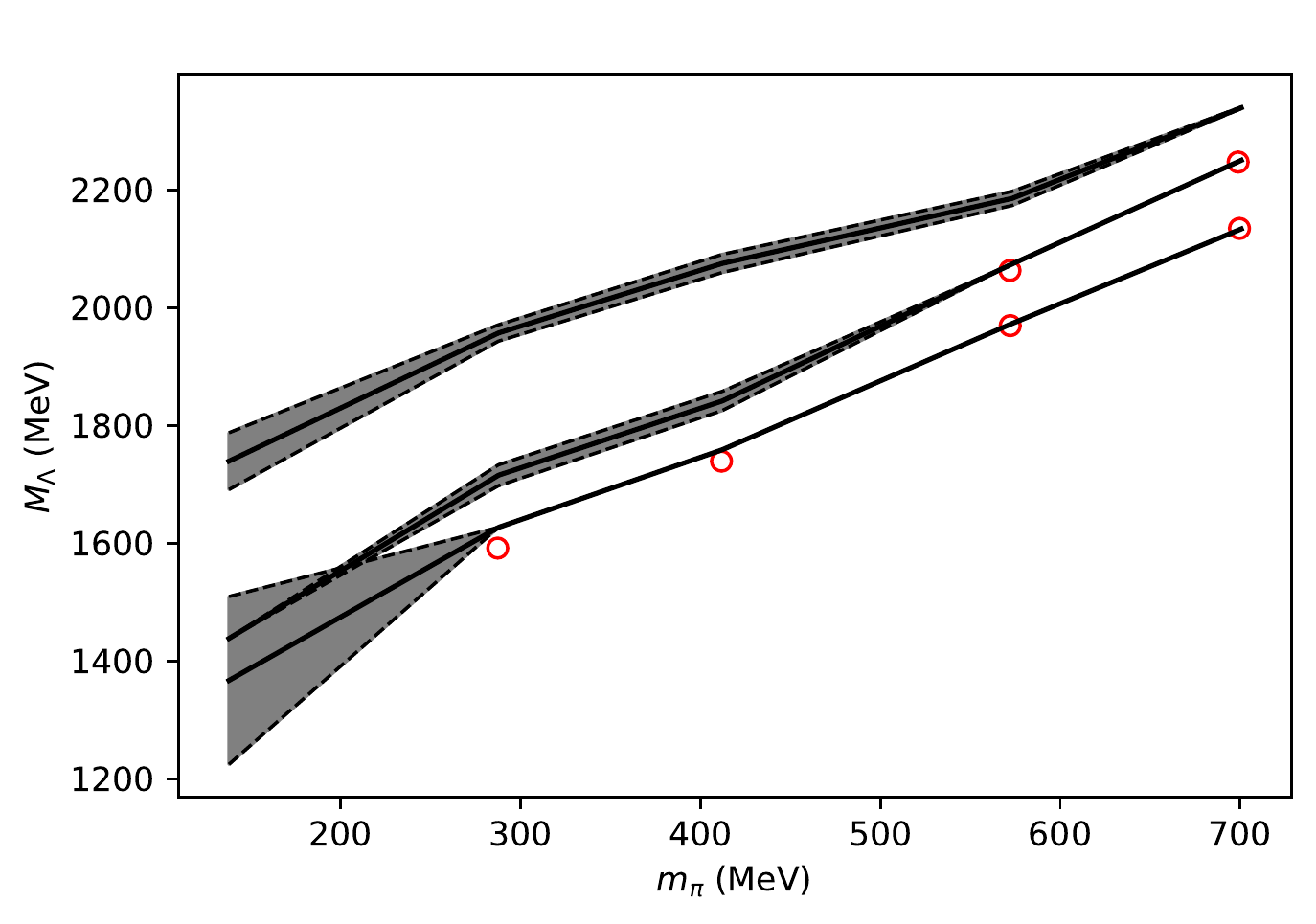}
  \includegraphics[scale=0.5,bb=0 0 402 266] {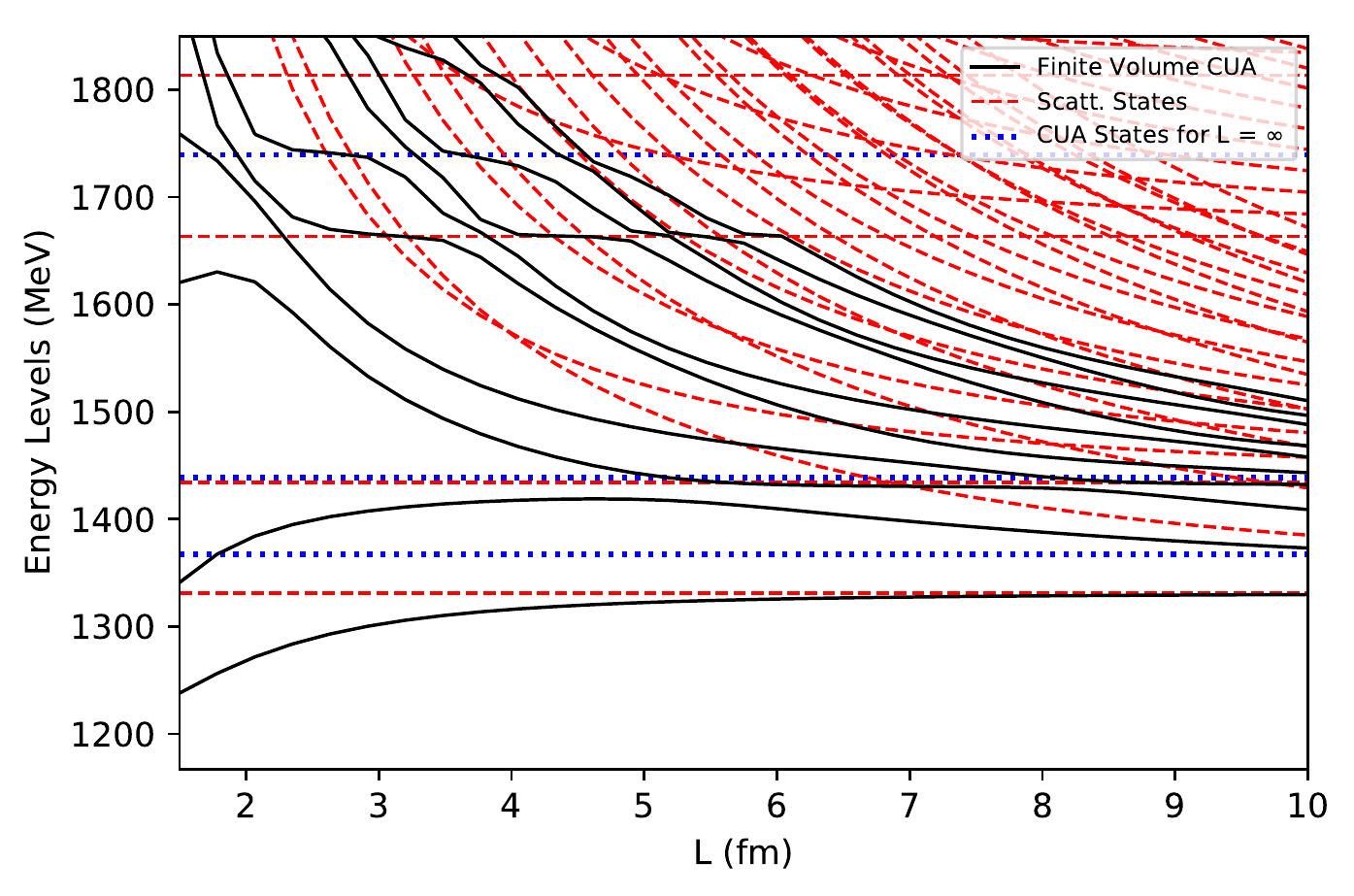}
\caption{\label{fig:finitejapan} Top plot: The $\Lambda$ baryon spectrum (resonances and bound states) obtained from infinite volume CUA. 
In case of bound states, the corresponding finite volume ($L=2.9$ fm) energy levels are shown as red circles. For resonances, 
the pole widths are shown as gray shaded areas. 
Bottom plot: Box size dependence of the full CUA energy levels (black solid lines) for physical hadron masses (set 1 in Table\,\ref{tab:input_values}). 
For comparison the coupled-channels non-interacting scattering states are shown as red 
dashed lines, while the positions of the CUA poles at $L=\infty$ are depicted as blue dotted lines.
}
\end{figure}

Employing the infinite volume chiral approach described in the previous section with the input of Table\,\ref{tab:input_values}, 
we obtain poles shown in the upper plot of Fig.\,\ref{fig:finitejapan} as functions of the pion mass. 
For pion masses above the physical one, the lowest or the two lowest poles are bound states. For these, we show their finite volume counterparts (for $L = 2.9\,$ fm) as red circles. 
As can be expected, the finite volume effects for bound states are relatively small and decrease with increasing pion masses. 
If the poles are resonances, their widths are shown as gray shaded areas. 

In the lower plot of Fig.\,\ref{fig:finitejapan}, we show the box size dependence of the finite volume CUA energy levels for 
the physical pion mass case (set 1 in Table \ref{tab:input_values}). Focussing on the region around $L \sim 3$ fm, which corresponds to the box size of the lattice 
QCD calculations considered in this work, one observes that the finite-volume energy levels (shown as black solid lines) in fact do not exhibit a clear signature 
of the two-pole pattern for the $\Lambda(1405)$. 
A similar finding was already discussed in the recent work of Tsuchida and Hyodo \cite{Tsuchida:2017gpb}, 
the conclusion of which is confirmed here. 
While the lowest black line can be interpreted as a $\pi \Sigma$ S-wave scattering state, the second energy level from below 
lies between the two $\Lambda(1405)$ poles at $L = \infty$ (lowest two blue dotted lines). 
The dynamics of this level should encode information on this double-pole structure. 
The third and fourth black lines display box-size dependences characteristic for scattering states, 
though they are strongly modified from the free scattering states (red dashed lines). 
This is caused by the strong and attractive $\bar{K}N$ interaction. 
These energy levels therefore likely also carry some information about the formation of the $\Lambda(1405)$ poles at infinite volume, 
although the 
relation is not straightforward and is obscured by the finite volume effects. The fifth energy level shows a plateau around $L \sim 3$ fm, 
related to the opening of the $\eta \Lambda$ channel, 
forming an S-wave scattering state. Finally, the sixth level displays another plateau, which 
is generated in conjunction with the $\Lambda(1670)$ resonance at finite volume. 

\begin{figure}
\centering
  \includegraphics[scale=0.5,bb=0 0 402 266] {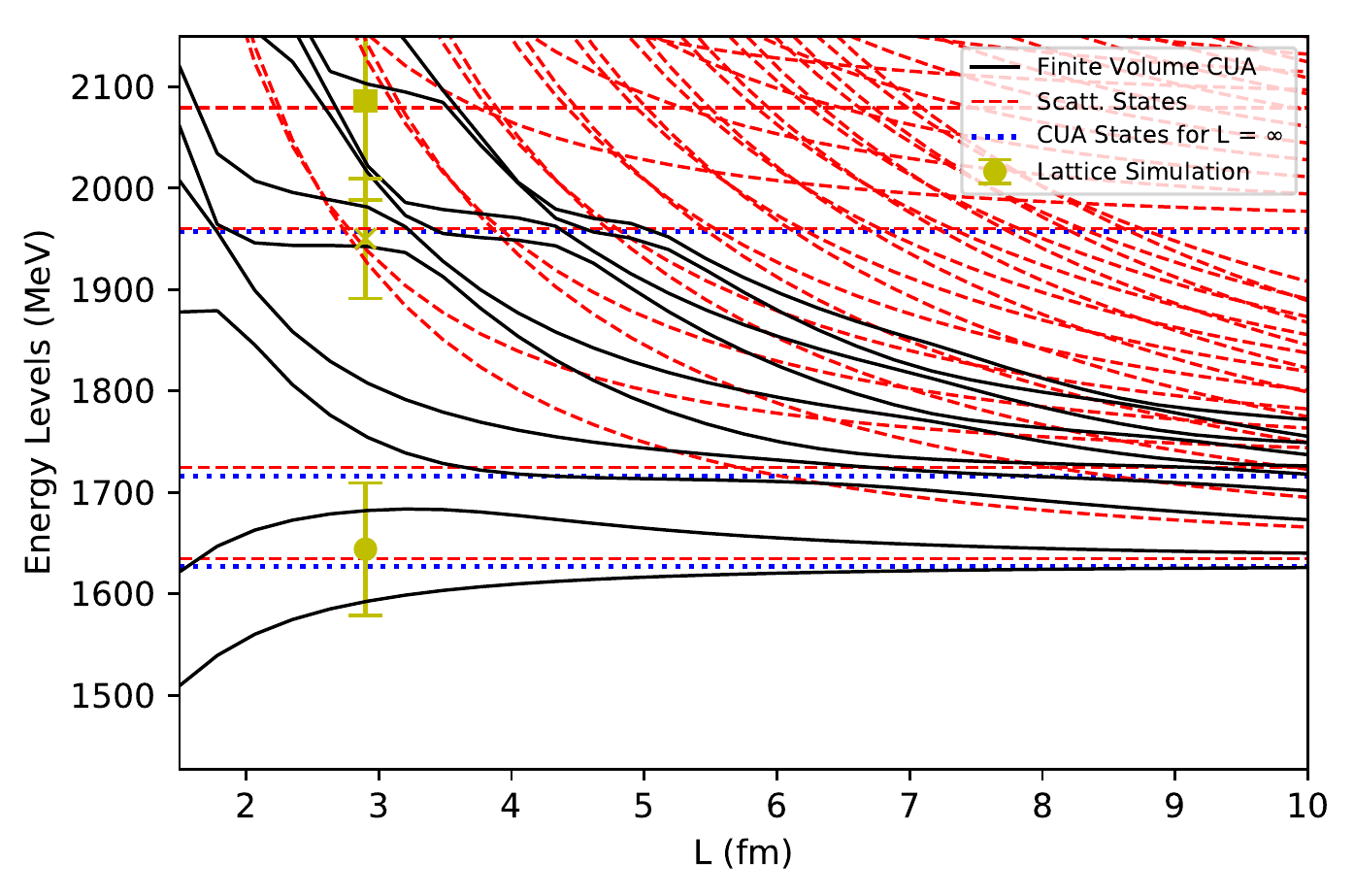}
\caption{\label{fig:L_dependence} Box size dependence of the full CUA energy levels (black solid lines) with hadronic input parameters of set 2
($m_{\pi} = 290$ MeV) in Table \ref{tab:input_values}. 
The corresponding lattice QCD results at $L=2.9$ fm are shown as greenish-yellow points. For comparison the 
coupled-channels non-interacting scattering states are shown as red 
dashed lines, while the positions of the CUA poles at $L=\infty$ are depicted as blue dotted lines.}
\end{figure}

Let us now compare the CUA results at finite volume with those obtained in lattice QCD. The results of Ref.\,\cite{Gubler:2016viv} at the 
lowest pion mass of $m_{\pi} = 290$ MeV are compared with the corresponding finite-volume CUA calculation in Fig.\,\ref{fig:L_dependence}. 
Note that the blue dotted lines in this figure, which show the CUA poles at $L = \infty$, correspond to the black lines in the upper plot of Fig.\,\ref{fig:finitejapan} 
at the corresponding pion mass. 
Compared with the physical pion mass case of the bottom plot in Fig.\,\ref{fig:finitejapan}, 
the lower $\Lambda(1405)$ pole at infinite volume becomes a bound state 
(compare the lowest red dashed and blue dotted lines). 
The roles of the two lowest energy-levels (black lines) are therefore interchanged: the lower line 
encodes information of 
the newly generated bound state corresponding to the 
lower $\Lambda(1405)$ pole at infinite box size, while the second level should be identified as a $\pi \Sigma$ scattering state. 
The third and fourth levels are again scattering states 
that are strongly modified due to the $\bar{K}N$ interaction, while the fifth and sixth 
ones show two plateaus that are very close in energy. 
One of them is 
an $\eta \Lambda$ S-wave scattering state and the other one (fifth) 
at $L \sim 3$ fm is likely related to the $\Lambda(1670)$ resonance, which has approached the $\eta \Lambda$ threshold from above. 
Comparing these finite volume CUA results with the lattice points (shown in greenish-yellow), it is observed that the lowest point 
obtained in the lattice QCD calculation of Ref.\,\cite{Gubler:2016viv} 
might be consistent with the $\pi \Sigma$ bound 
state influenced by the $\bar{K} N$-coupled-channels dynamics, while the second one agrees well with the energy level that reflects the infinite volume $\Lambda(1670)$. 
The physical meaning of the third lattice QCD state is less clear, 
as CUA generates only scattering states in the region of this energy level (see Fig.\,2). 
One hence cannot make any statement about its nature from the comparison between CUA and lattice QCD data. 
It is apparently a state 
with a structure that goes beyond the physics of meson-baryon states described by CUA and is most likely a reflection of 
the experimentally observed infinite-volume $\Lambda(1800)$. 
 
\begin{figure}
\centering
  \includegraphics[scale=0.5,bb=0 0 402 262] {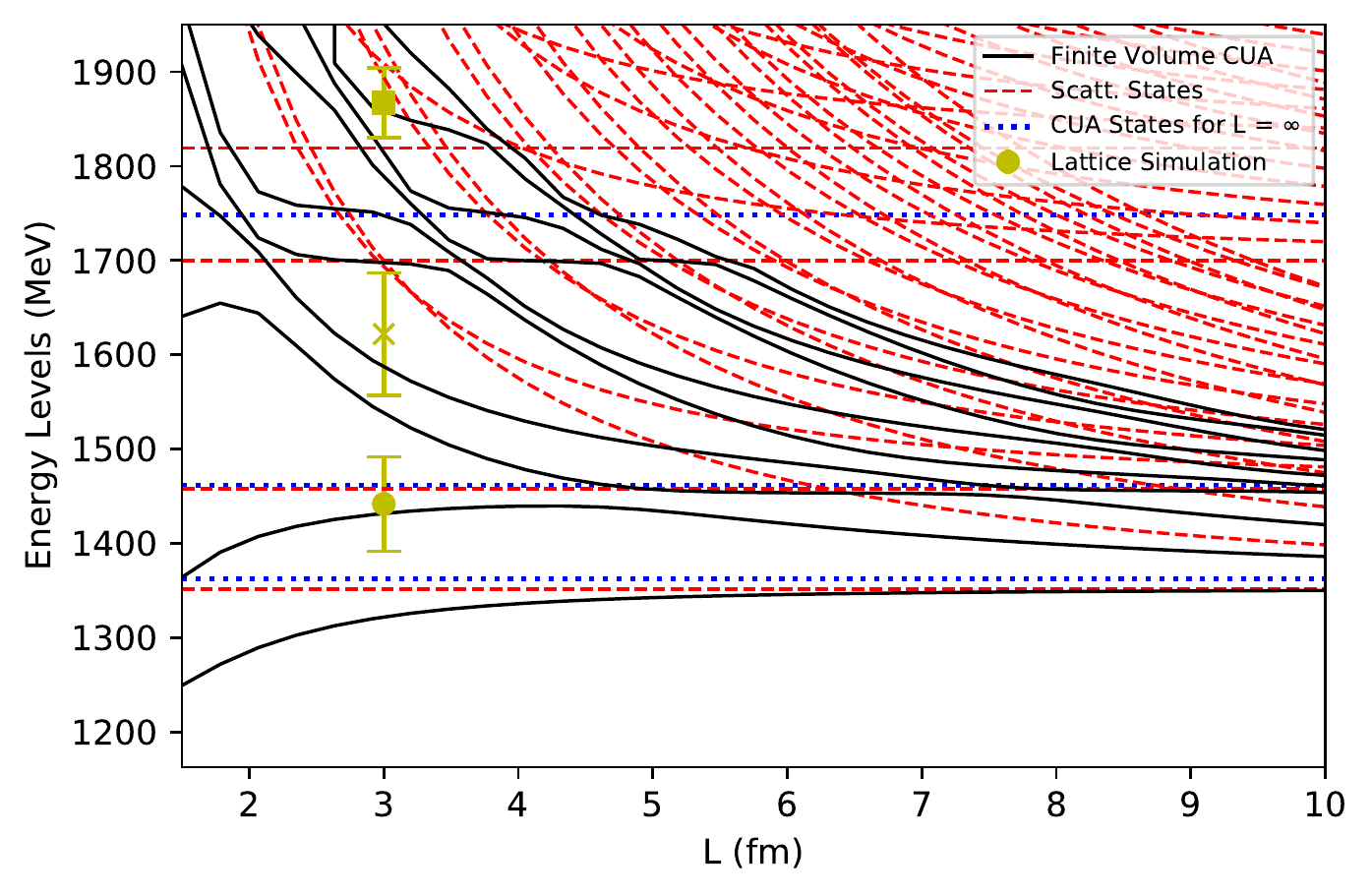}
\caption{\label{fig:L_dependence_Adelaide} Same as Fig.\,\ref{fig:L_dependence}, but for the results of Ref.\,\cite{Menadue:2018abc} with $m_{\pi} = 152$ MeV. 
See text for more details.}
\end{figure}

Next, we compare the recent lattice results of the Adelaide group \cite{Menadue:2013xqa,Hall:2014uca,Hall:2014gqa,Liu:2016wxq,Menadue:2018abc} 
to our finite volume CUA calculation. For our comparison, we employ the data reported in the most recent Ref.\,\cite{Menadue:2018abc}, 
which also provides the most complete spectrum of ground and excited states. The corresponding values for baryon masses, meson masses and decay 
constants, used as input for the CUA calculation, are given in Table II of Ref.\,\cite{Molina:2015uqp}. 
The comparison is shown in Fig.\,\ref{fig:L_dependence_Adelaide} for the lowest pion mass, $m_{\pi} = 152$ MeV. 
One notes that 
the lowest finite-volume CUA eigenstate corresponding to a $\pi \Sigma$ scattering state is missed. 
However, this lattice simulation well reproduces the 
energy level located between the two infinite volume $\Lambda(1405)$ poles. 
The next lattice eigenstate lies close to the third and fourth CUA levels, which, as already mentioned in the discussion of the lower plot of 
Fig.\,\ref{fig:finitejapan}, behave like scattering states, but are strongly modified by the attractive $\bar{K}N$ interaction. 
We speculate that the second lattice eigenstate corresponds to these
scattering states. They could (at least, partially) be extracted in Ref.\,\cite{Menadue:2018abc} 
as they have sufficient overlap with the three quark operator
due to the strong $\bar{K}N$ attraction.
This speculation will be further
confirmed in the comparison between the lattice data of ours and the
Adelaide group for heavier quark masses. 
The assignement of the third lattice state is less clear. It may be identified with 
either the finite-volume energy level corresponding to the $\Lambda(1670)$ or, more likely, with that of the physical $\Lambda(1800)$ (which is however 
not obtained in the CUA spectrum).

Let us here briefly discuss the differences between the lattice QCD results of Refs.\,\cite{Gubler:2016viv} and \cite{Menadue:2018abc} 
(and the other recent works \cite{Menadue:2013xqa,Hall:2014uca,Hall:2014gqa,Liu:2016wxq} from the Adelaide group). 
The analyses for both of them were done using the same gauge configurations of the PACS-CS Collaboration \cite{Aoki:2008sm} and 
employ multiple local three-quark operators to generate the $\Lambda$ baryon states. 
They, however, differ in a number of aspects, the main differences being the following: 
1) Ref.\,\cite{Menadue:2018abc} partially quenched the strange valence quarks appearing 
in its analysis (i.e. the mass of the strange valence quark was modified from the 
value used to generate the PACS-CS gauge configurations), such that the simulated kaon 
mass at the physical point better reproduces the experimental value. This was not done 
in Ref.\,\cite{Gubler:2016viv}. 
2) While Ref.\,\cite{Menadue:2018abc} employed the Sommer scale \cite{Sommer:1993ce} to determine the lattice spacing $a$ individually at each 
pion mass, Ref.\,\cite{Gubler:2016viv} used the value provided in Ref.\,\cite{Aoki:2008sm} and applied it to all pion mass cases. 
3) Ref.\,\cite{Menadue:2018abc} used operators with different degrees of smearing to 
increase the number of bases in its variational analysis, while Ref.\,\cite{Gubler:2016viv} only used operators with different Dirac-structures. 

\begin{figure}
\centering
  \includegraphics[scale=0.68,bb=0 0 360 216] {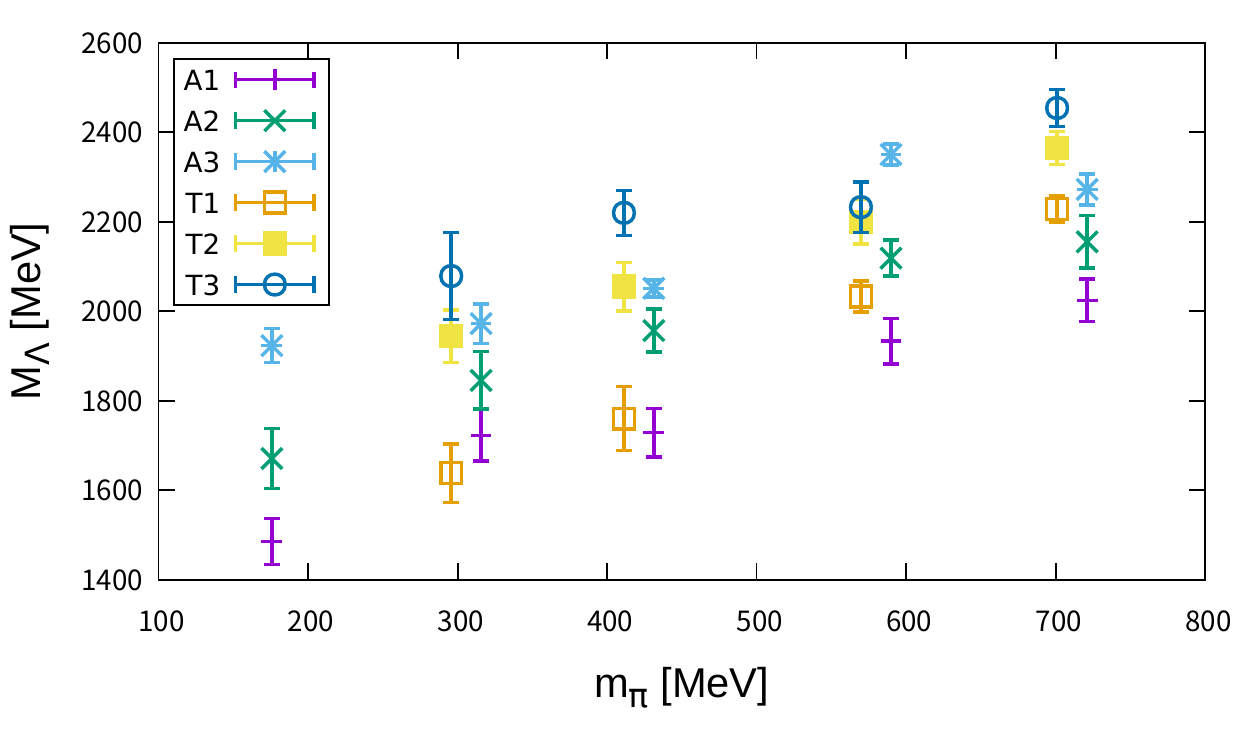}
\caption{\label{fig:Lattice_comp} Comparison of the lattice QCD spectra of Refs.\,\cite{Menadue:2018abc} (A1-3, cross symbols) and 
\cite{Gubler:2016viv} (T1-3, circle and square symbols). 
The data of Ref.\,\cite{Menadue:2018abc} are rescaled to match the scale setting prescription of Ref.\,\cite{Gubler:2016viv} and are 
horizontally shifted to improve the visibility of the plot.
}
\end{figure}

The effect of point 2) can be easily eliminated by a rescaling of the respective results. 
To get an idea on potential effects of the other two, we compare the rescaled spectra in Fig.\,\ref{fig:Lattice_comp}.
Comparing the results for the pion masses $m_{\pi} \sim 290$ MeV and $m_{\pi} \sim 410$ MeV, it is seen that the ground states 
agree within errors. For the excited states, Ref.\,\cite{Menadue:2018abc} seems to be able to extract one state that was missed 
in Ref.\,\cite{Gubler:2016viv}, while the third state of Ref.\,\cite{Menadue:2018abc} again agrees well with the second state of Ref.\,\cite{Gubler:2016viv}. 
Even though the precision of these lattice results is certainly not good enough to make any definite statements, the additional second state of 
Ref.\,\cite{Menadue:2018abc} may be most naturally explained by the larger class of operators used in their analysis, which made 
it possible to probe states with a larger variety of spatial structures.
 
As can be appreciated from the above comparison between finite volume CUA and lattice QCD spectra in Figs.\,\ref{fig:L_dependence} and \ref{fig:L_dependence_Adelaide}, 
it is generally difficult to extract genuine scattering states in lattice QCD calculations employing 
only local three-quark interpolating fields with small overlap with highly non-local meson-baryon (hence five-quark) scattering states, 
as it is the case in Refs.\,\cite{Gubler:2016viv,Menadue:2013xqa,Hall:2014uca,Hall:2014gqa,Liu:2016wxq,Menadue:2018abc} 
discussed in this paper. 
To generate such scattering states, the use of five-quark operators will therefore likely be necessary in future lattice studies of this channel.  

Furthermore, it is clear from the finite volume CUA spectrum and the computed lattice QCD energy-levels that a 
direct identification of the finite-volume spectrum with the two $\Lambda(1405)$ poles at infinite volume is not possible. 
To unambiguously establish the two-pole nature of the $\Lambda(1405)$ from a first-principle lattice QCD calculation, one would therefore 
likely need a much larger number of very precise lattice QCD data points at multiple volumes as large as $L\sim 7$ fm to, for instance, perform an analysis as proposed in 
Ref.\,\cite{MartinezTorres:2012yi}.

\subsection{Flavor structure}
\begin{figure}[t!] 
\center
\includegraphics[width=0.9\linewidth,bb=0 0 390 266]{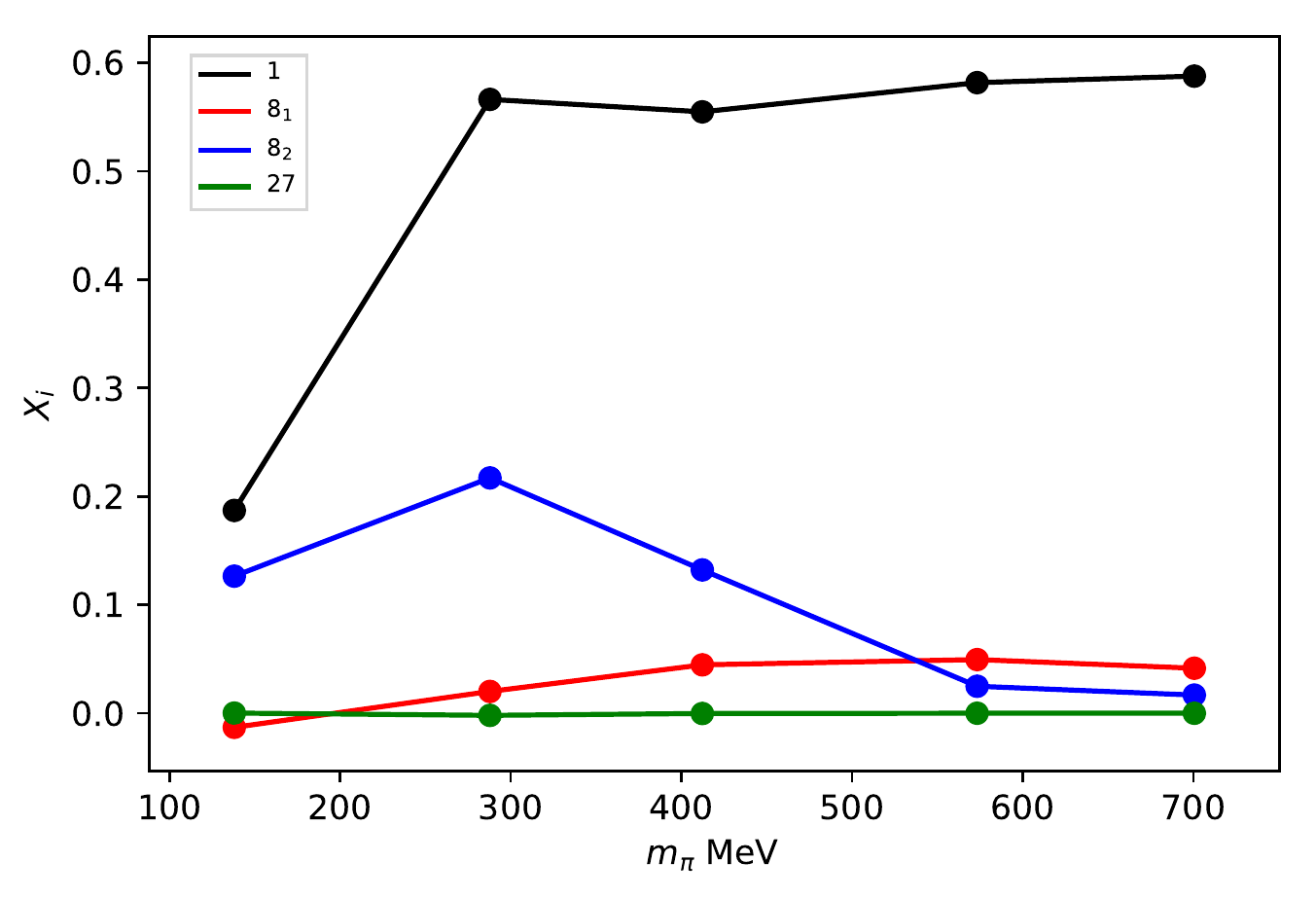}
\includegraphics[width=0.9\linewidth,bb=0 0 390 266]{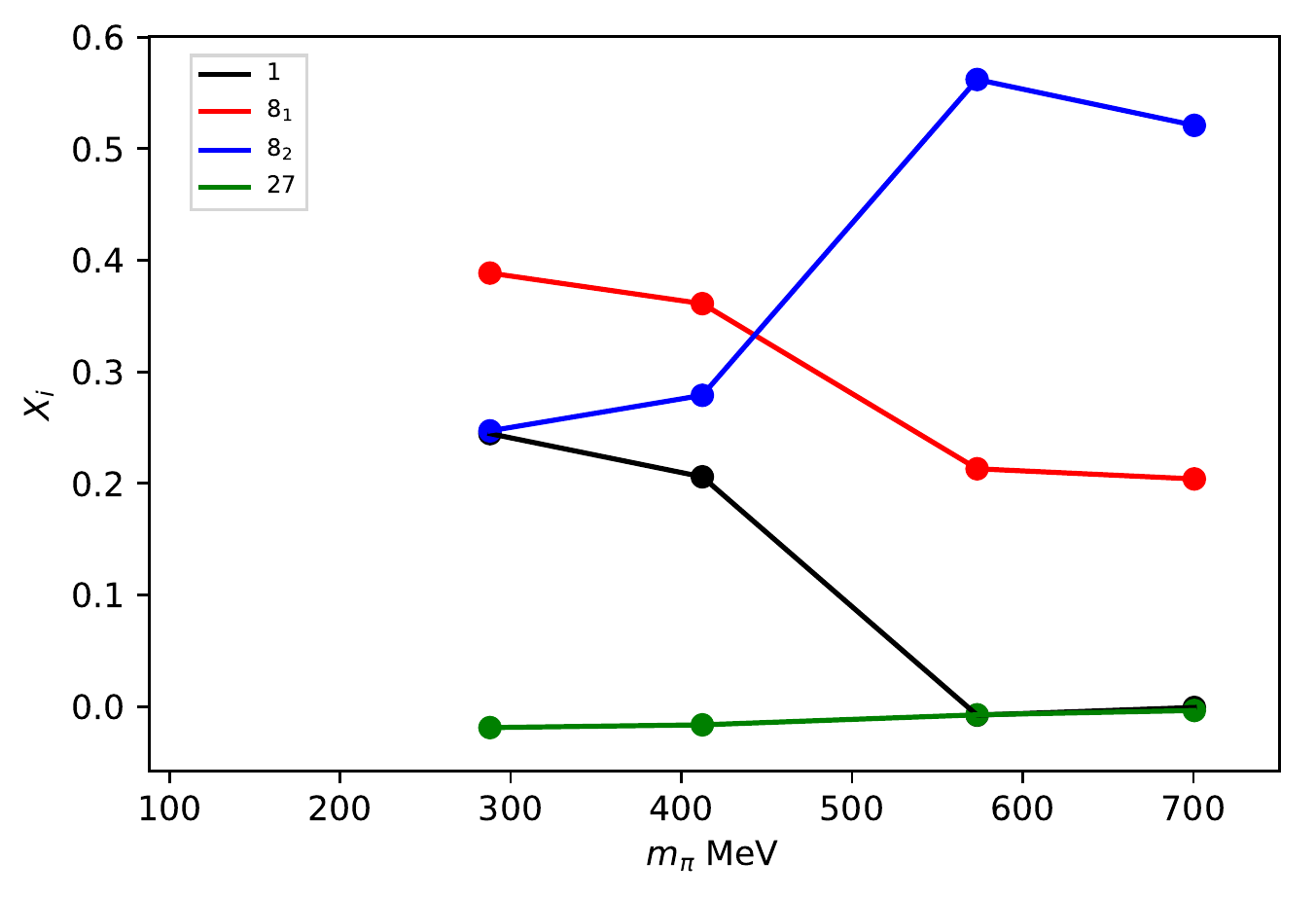}
\includegraphics[width=0.9\linewidth,bb=0 0 390 266]{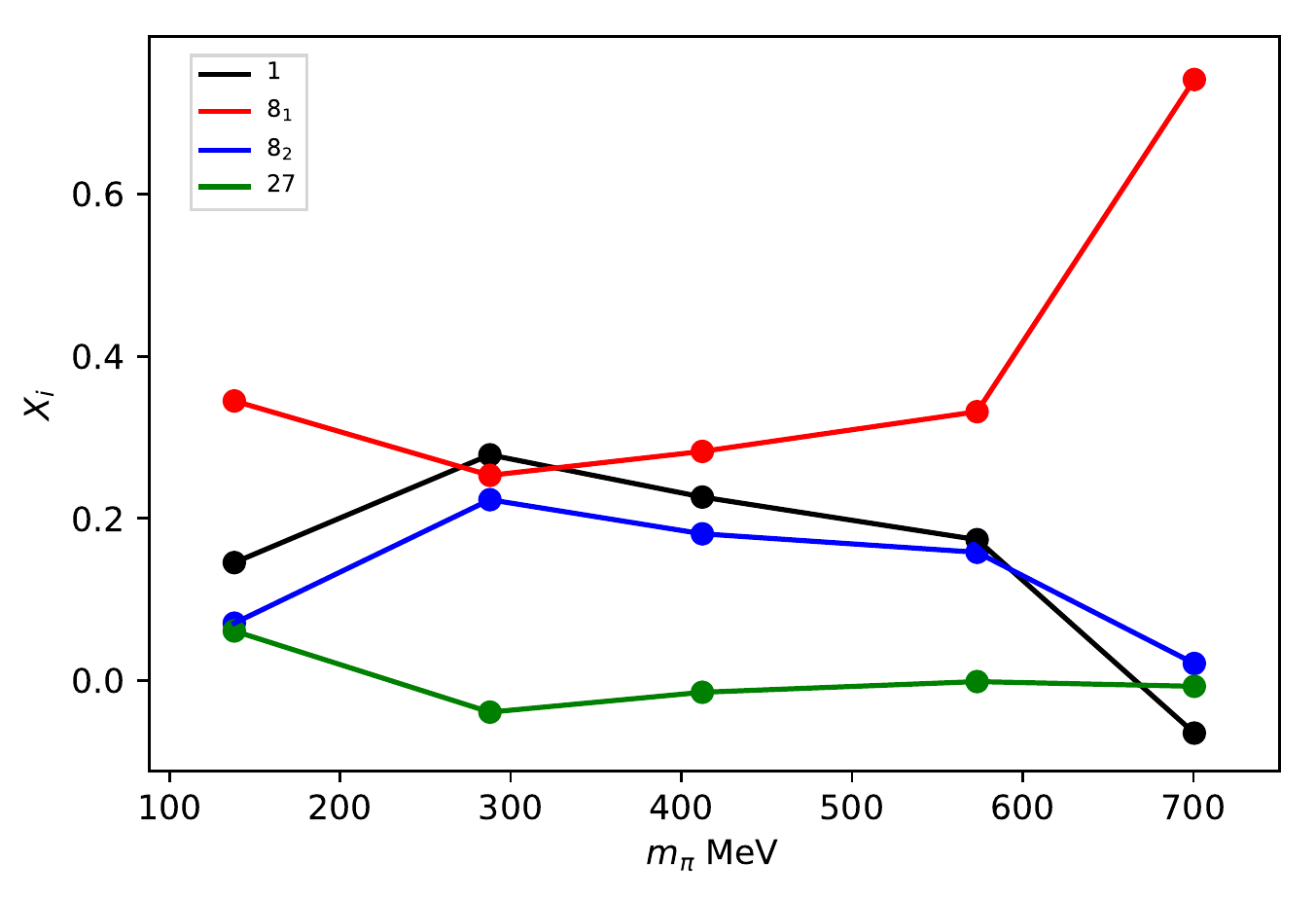}
\caption{Compositeness of Eq.\,(\ref{eq:comp}) for the CUA states at infinite box size, with respect to the $SU(3)$ basis. The two top plots 
show the decomposition of the two $\Lambda(1405)$ poles, the bottom one that for the $\Lambda(1670)$.} \label{fig:compsu3}
\end{figure}

%
%
%
%
With the compositeness measure discussed in Section \ref{sec:Compo}, 
we can calculate the contributions of the different meson-baryon channels, in both [flavor $SU(3)$ or particle] representations, to each $\Lambda$ state. 
Our infinite-volume CUA results for the flavor $SU(3)$ basis are plotted in Fig.\,\ref{fig:compsu3}, 
%
%
%
\begin{figure}[t!] 
\center
\includegraphics[width=0.9\linewidth,bb=0 0 390 266]{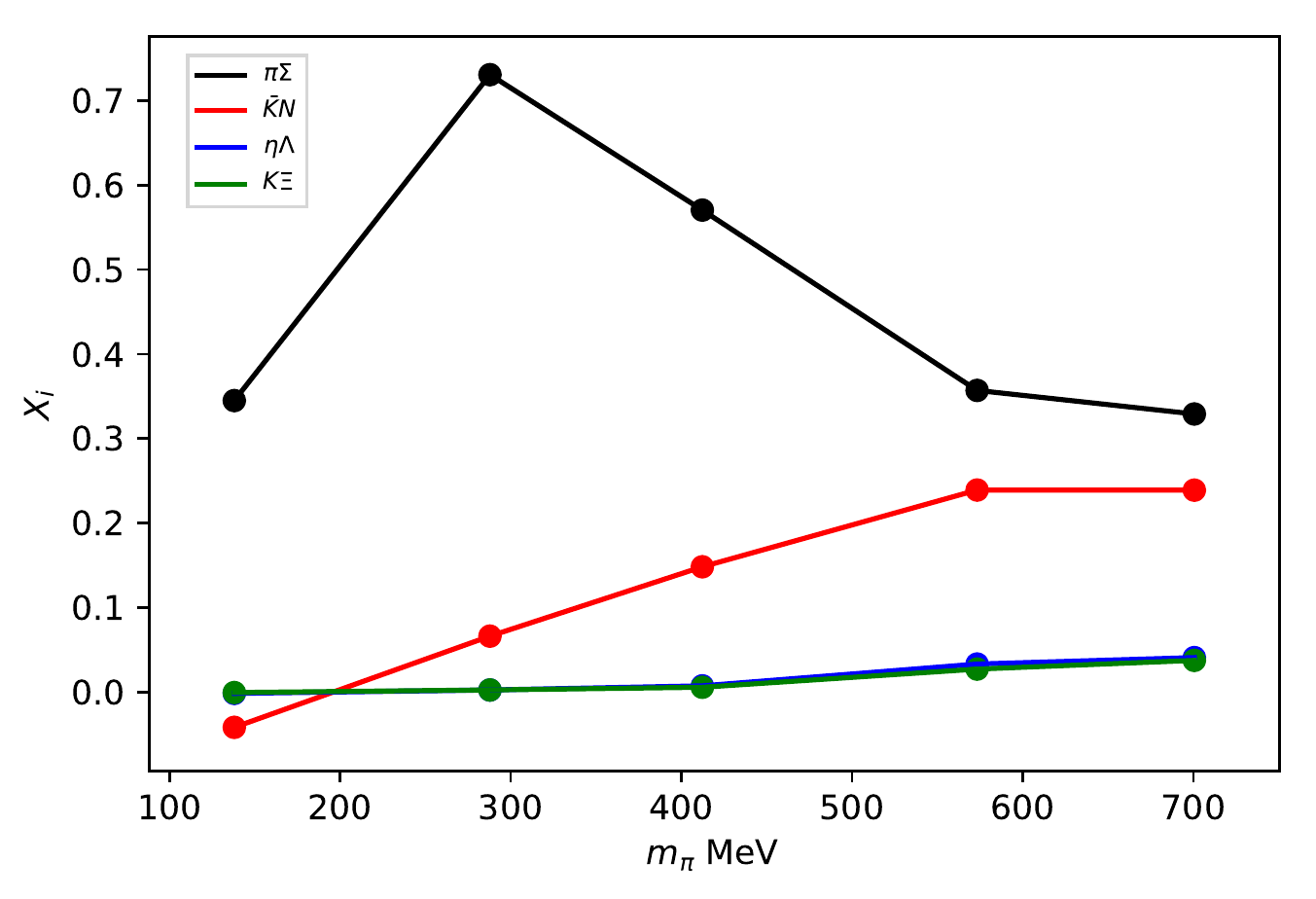}
\includegraphics[width=0.9\linewidth,bb=0 0 390 266]{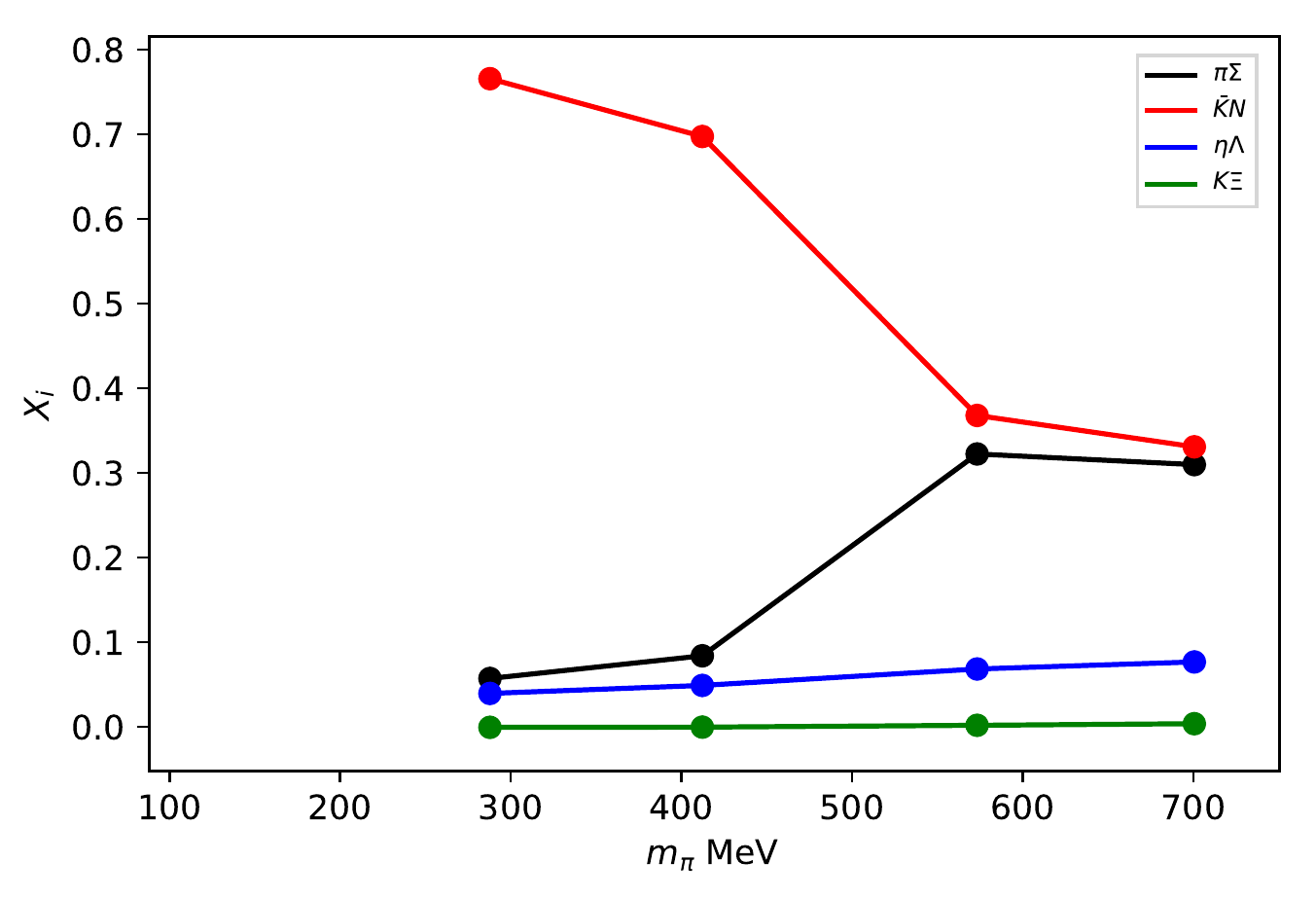}
\includegraphics[width=0.9\linewidth,bb=0 0 390 266]{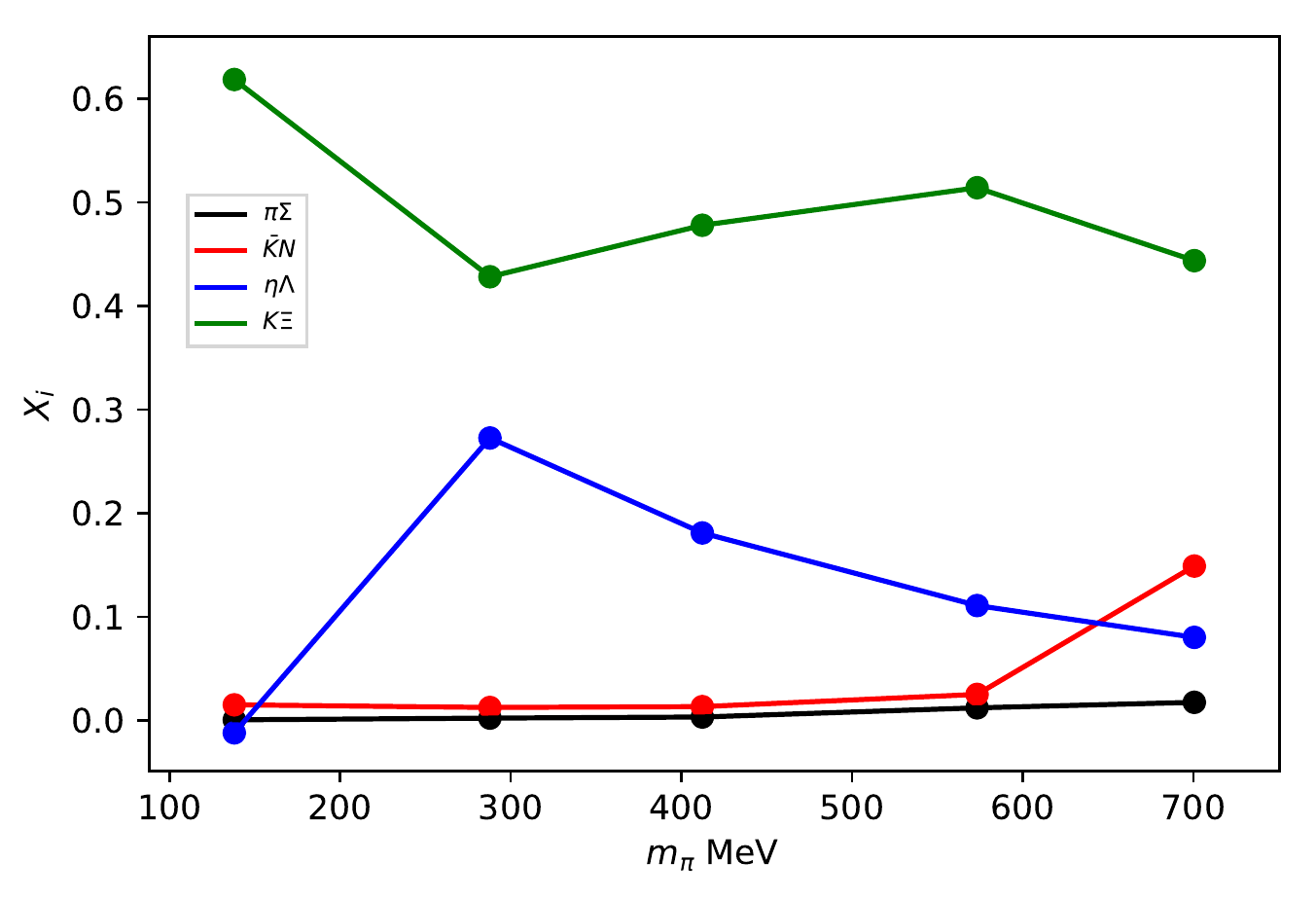}
\caption{Compositeness of Eq.\,(\ref{eq:comp}) for the CUA states at infinite box size, with respect to the particle basis. The two top plots 
show the decomposition of the two $\Lambda(1405)$ poles, the bottom one that for the $\Lambda(1670)$.} \label{fig:compphys}
\end{figure}
%
from which it is understood that the lower $\Lambda(1405)$ pole within this effective approach is dominated by the singlet contribution, with 
a small and decreasing octet component with increasing pion mass. The second $\Lambda(1405)$ pole, on the other hand, is dominated by 
the two octet components, however with a sizable singlet contribution at intermediate pion masses. Note that this pole becomes virtual in our 
framework for the lowest pion mass, which is why the 
corresponding points are missing. The $\Lambda(1670)$ state is for most pion masses 
dominated by the $|\mathbf{8_1}\rangle$, with sizable contributions of both $|\mathbf{1}\rangle$ and $|\mathbf{8_2}\rangle$ at 
intermediate $m_{\pi}$. The $|\mathbf{27}\rangle$ plays no major role for any of the states discussed in this work. 
All these results agree with those reported in Ref.\,\cite{Jido:2003cb}. 

Looking at the same states in the particle basis in Fig.\,\ref{fig:compphys}, one recovers the well established fact (see, for instance the 
review in Ref.\,\cite{Tanabashi:2018oca}) that the lower $\Lambda(1405)$ state is predominantly a $\pi \Sigma$ molecule, while the upper one 
is dominated by $\bar{K}N$. The $\Lambda(1670)$, conversely, is dominated by the $K \Xi$ molecular component.  

In the lattice QCD study of Ref.\,\cite{Gubler:2016viv}, the various states were similarly decomposed into flavor $SU(3)$ representations according to 
Eqs.\,(\ref{eq:su3_latt_1}) and (\ref{eq:su3_latt_2}). Referring to Ref.\,\cite{Gubler:2016viv} for details, we here reproduce the relevant 
$g_{i}^{\bm{1}}$ and $g_{i}^{\bm{8}}$ normalized singlet and octet overlaps from that work in Table\,\ref{tab:g_values}. 
\begin{table*}
\renewcommand{\arraystretch}{1.2}
\begin{center}
\caption{Values of $g_{n}^{\bm{1}}$ and $g_{n}^{\bm{8}}$, defined in Eqs.\,(\ref{eq:su3_latt_1}) and (\ref{eq:su3_latt_2}), obtained in the lattice QCD study 
of Ref.\,\cite{Gubler:2016viv}. $n=1$, 2, 3 specifies the three lowest energy-states obtained in that work.}
\begin{tabular}{ccc|cc|cc}  
\toprule
set & $g_{1}^{\bm{1}}$ & $g_{1}^{\bm{8}}$ & $g_{2}^{\bm{1}}$ & $g_{2}^{\bm{8}}$ & $g_{3}^{\bm{1}}$ & $g_{3}^{\bm{8}}$ \\ \midrule
2 &  0.944(15) & 0.056(11) & 0.013(6) & 0.987(125) & 0.011(3) & 0.989(93) \\
3 & 0.941(13) & 0.059(10) & 0.013(7) & 0.987(209) & 0.014(6) & 0.986(129) \\
4 & 0.963(8) & 0.037(7) & 0.007(3) & 0.993(235) & 0.001(2) & 0.999(358) \\
5 & 0.984(3) & 0.016(2) & 0.005(2) & 0.995(212) & 0.001(1) & 0.999(127) \\
\bottomrule
\label{tab:g_values}
\end{tabular}
\end{center}
\end{table*}
We see that the lowest 
state extracted on the lattice was a singlet dominated state with a small octet contribution of about 5\% at $m_{\pi} = 288\,\mathrm{MeV}$ (set 2 in Table\,\ref{tab:g_values}), 
which decreases further with increasing pion mass. The second state [which we identify with the $\Lambda(1670)$ in this work], 
is octet dominated with a singlet contribution of at most 1\,\%. Note that octet components 
$|\mathbf{8_1}\rangle$ and $|\mathbf{8_2}\rangle$ cannot be distinguished on the lattice. 
The definitions of the $g$-values of Eqs.\,(\ref{eq:su3_latt_1}) and (\ref{eq:su3_latt_2}) and the compositeness measure of Eq.\,(\ref{eq:comp}) are furthermore 
not equivalent. A comparison between the two approaches is therefore only possible in a qualitative manner. 

Making such a comparison, one finds that, because the lowest lattice QCD state is predominantly a singlet state, it can most likely be identified 
with the corresponding lowest CUA state, which except for the lowest (physical) pion mass, not studied in Ref.\,\cite{Gubler:2016viv}, 
is mostly a singlet as well. 
As depicted in the top plot of Fig.\,\ref{fig:compphys}, the dominant component of this state within the CUA approach adopted in this work\footnote{The importance of 
other components, as $\bar K N$ appreciably depends on the pion mass, or on some specific details (for example, the renormalization procedure) of the employed CUA 
(see for instance the differences between results given in Tables 1 and 3 of  Ref.\,\cite{Jido:2003cb}.)} is $| \pi \Sigma \rangle$. 
Let us briefly discuss here how this finding compares with the results of 
Ref.\,\cite{Hall:2014uca}, where the light ($u$ and $d$) and strange quark contributions to the Sachs magnetic form factor were computed 
to determine the flavor components of the lowest lattice energy-level. 
For pion masses corresponding to sets 2 $-$ 5 in Table \ref{tab:input_values}, the contributions of the light and strange quarks to the form factor are 
of comparable magnitude, which is consistent with our conclusion that this state is dominated by the $| \pi \Sigma \rangle$ component. 
As the pion mass is further reduced to the (almost) physical value of set 1, the strange quark contributions to the Sachs magnetic form factor 
suddenly decreases to zero, indicating that the corresponding state is mainly a $|\bar{K} N \rangle$ state. While we cannot independently 
confirm this finding due to our lack of lattice data for set 1, we do see some hint of a sudden structural change when approaching the 
lowest pion mass both in Figs.\,\ref{fig:compsu3} and \ref{fig:compphys}. 

The identification of the second lattice QCD state 
with the $\Lambda(1670)$ obtained in the CUA can be also qualitatively confirmed by comparing their flavor components, 
which are both octet-dominated. 
The agreement is, however, not perfect here because of the sub-dominant but nevertheless non-negligible singlet contribution to the CUA state 
and its relatively large pion mass dependence, 
as seen in the bottom plot of Fig.\,\ref{fig:compsu3}. Its lattice counterpart meanwhile is an almost completely pure octet state 
and does hardly change with $m_{\pi}$. 
%

\section{Summary and conclusions \label{sec:SumCon}}
We have studied the spectrum and flavor structure of the negative-parity spin-1/2 $\Lambda$ baryons by comparing the 
findings from two first principle lattice QCD calculations and the CUA based on effective hadronic degrees of freedom. 
To make a direct and realistic comparison between the two approaches possible, we have modified the CUA by using non-physical hadron  
masses and meson decay constants, such that they correspond to the situation studied on the lattice. 
We furthermore investigated the $\Lambda$ baryon spectrum in the CUA both at infinite and finite volume, which enables us to 
examine and clarify potential effects of the finite lattice size in the lattice QCD calculation. 
Besides studying the $\Lambda$ baryon energy spectrum, we also compared the flavor structure of the obtained states, thus 
providing additional information for determining which lattice QCD state can be identified with which CUA state. 

The main findings of this work can be summarized as follows. 
\begin{itemize}
\item
While the CUA at infinite volume generates two poles corresponding to the $\Lambda(1405)$ and one corresponding to the $\Lambda(1670)$, 
finite volume
lattice QCD simulations find only one state 
that clearly encodes information about the $\Lambda(1405)$ at infinite volume and, at least for 
Ref.\,\cite{Gubler:2016viv}, another one representing the $\Lambda(1670)$. 
As it was already discussed 
in Ref.\,\cite{Tsuchida:2017gpb} and again confirmed in this work, the fact that only one state closely related to the $\Lambda(1405)$ 
is found is actually a finite-volume effect and it is not related to any other deficiency of the lattice QCD calculation. 
Let us furthermore stress that there is generally no unique and direct one-to-one conrrespondence between finite volume 
energy levels and infinite volume resonances and information about any specific infinite volume resonance hence can be spred 
over multiple finite volume energy levels. This phenomenon is for instance observed in the third and fourth energy levels of Fig.\,\ref{fig:L_dependence}, 
which behave like scattering states, but are clearly modified by the strong and attractive $\bar{K}N$ interaction and therefore carry information about 
a potential $\bar{K}N$ resonance at infinite volume.

\item
The lattice QCD studies of Refs.\,\cite{Gubler:2016viv,Menadue:2013xqa,Hall:2014uca,Hall:2014gqa,Liu:2016wxq,Menadue:2018abc}, 
which use three-quark interpolating fields to generate the physical states, 
obtain only a limited number of energy levels compared to the finite volume CUA (see Figs.\,\ref{fig:L_dependence} and \ref{fig:L_dependence_Adelaide}). 
Especially, energy levels that at a specific box size closely resemble free meson-baryon scattering states, are mostly missed in the 
lattice calculations available so far. 
Other energy-levels that are strongly influenced by the formation of bound states or resonances, 
are partially reproduced by the lattice QCD calculations, even though better precision will be needed 
to make a more thourough comparison between lattice QCD and CUA. 
It can be expected that the use of a richer variety of operators on the lattice (such as local or non-local five-quark operators), will 
be needed to generate the scattering states and hence the full spectrum at any given lattice size. 
Such a study will also be needed to provide compelling evidence for the two-pole nature of the 
$\Lambda(1405)$ from lattice QCD.

\item
A qualitative comparison between the flavor $SU(3)$ decompositions of respective lattice QCD and CUA energy levels
suggests that the lowest state generated on the lattice can be identified with the lower CUA $\Lambda(1405)$ 
pole, which is mostly a $\pi \Sigma$ molecule. This conclusion is, however, only valid for pion masses of about 290 MeV or larger. 
As the pion mass is lowered to its physical value, there are indications (see Ref.\,\cite{Hall:2014uca}) that the flavor composition of this $\Lambda(1405)$ state is 
strongly modified.
\end{itemize}

To conclude, let us discuss the remaining obstacles for reaching a complete understanding  
of the spectrum of the negative-parity spin-1/2 $\Lambda$ baryons and especially to disentangle the intricate dynamics of 
its lowest member, the $\Lambda(1405)$ from QCD. 
Since for physical masses, all negative parity 
$\Lambda$ baryons are resonances, the goal of any theoretical effort to describe them will boil down to the calculation of the 
complex scattering amplitudes of the relevant decay channels and the location of any poles on them. 
On the lattice, this can presently only be done by performing a calculation of multiple energy levels of interacting two-particle states on multiple volumes, 
by which one aims to constrain the scattering amplitude and its energy dependence stringently enough to locate the 
pole positions with reasonable precision. 

Considerable efforts have been devoted to this line of work in recent years, especially by connecting experimental infinite volume 
scattering amplitudes with their finite volume lattice counterparts by making use of CUA (as done in this work) or other types of 
hadronic effective field theories (see Refs.\,\cite{Briceno:2017max,Mai:2021lwb} for recent reviews and references therein, 
but also Ref.\,\cite{Aoki:2020bew} for an alternative approach), and much progress has been made. Neverthess, 
conlusive analyses of scattering amplitudes based on lattice QCD data have so far 
only been achieved for meson-meson 
scattering channels, while computations of meson-baryon states 
at the physical point are still challenging. 
For the negative parity $\Lambda$ baryons, the situation is made even more difficult due the existence of annihilation diagrams, 
which are noisy and computationally costly. The confirmation (or rebuttal) and clarification of the two-pole nature of 
the $\Lambda(1405)$ therefore remains to be a formidable task, yet to be overcome by the lattice QCD community.

\begin{acknowledgments}
P.G. is supported by 
the Grant-in-Aid for Early-Carrier Scientists (JSPS KAKENHI Grant Number JP18K13542), 
Grant-in-Aid for Scientific Research (C) 
(JSPS KAKENHI Grant Number JP20K03940) and 
the Leading Initiative
for Excellent Young Researchers (LEADER) of the Japan
Society for the Promotion of Science (JSPS). 
M.O. is supported by JSPS KAKENHI Grant Number JP19H05159. 
M.O. and P.G. are also supported by JSPS  KAKENHI Grant Number  20K03959. 
This research has been supported
by the Spanish Ministerio de Econom\'{i}a y Competitividad (MINECO) and the European Regional Development Fund
(ERDF) under contracts FIS2017-84038-C2-1-P and by Generalitat
Valenciana under contract PROMETEO/2020/023 and by the EU STRONG-2020 project under the program H2020-INFRAIA-2018-1, 
grant agreement No. 824093.
\end{acknowledgments}

\bibliography{references_PG}

%
%
%

%
\end{document}